\documentclass[conference]{IEEEtran}
\input{psfig.sty}
\usepackage{subfigure}
\usepackage{multirow}
\usepackage{psfig}
\usepackage{graphicx}
\usepackage{amsmath}
\usepackage{epsfig}
\usepackage{amsfonts}
\usepackage{amssymb}
\usepackage{amsthm}
\usepackage{algorithm}
\usepackage{algorithmic}
\usepackage[usenames,dvipsnames]{pstricks}
\usepackage{arydshln}

\newcommand{\argmin}{\mathop{\text{argmin}}}

\newcommand{\vv}{{\bf v}}
\newcommand{\vs}{{\bf s}}
\newcommand{\vx}{{\bf x}}
\newcommand{\vy}{{\bf y}}
\newcommand{\vz}{{\bf z}}
\newcommand{\vn}{{\bf n}}

\newcommand{\pp}{{\bf P}}
\newcommand{\mh}{{\bf H}}
\newcommand{\mi}{{\bf I}}
\newcommand{\mg}{{\bf G}}
\newcommand{\mf}{{\bf F}}
\newcommand{\ma}{{\bf A}}
\newcommand{\mb}{{\bf B}}
\newcommand{\md}{{\bf D}}

\newcommand{\sm}{{\mathbb S}_{n_t,{\mathbb A}}}
\newcommand{\sa}{{\mathbb A}}

\begin{document}
\twocolumn

\title{{\huge 
Pseudo-random Phase Precoded Spatial Modulation
}}

\author{T. Lakshmi Narasimhan, Yalagala Naresh, Tanumay Datta$^{\dagger}$, 
A. Chockalingam \\
Department of ECE, Indian Institute of Science, Bangalore 560012, India \\ 
$\dagger$ \ Presently with Broadcom India Research Private Limited, 
Bangalore 560103, India}

\IEEEaftertitletext{\vspace{-0.6\baselineskip}}
\maketitle
\begin{abstract}
Spatial modulation (SM) is a transmission scheme that uses multiple transmit 
antennas but only one transmit RF chain. At each time instant, only one among 
the transmit antennas will be active and the others remain silent. The 
index of the active transmit antenna will also convey information bits in 
addition to the information bits conveyed through modulation symbols (e.g., 
QAM). Pseudo-random phase precoding (PRPP) is a technique that can achieve 
high diversity orders even in single antenna systems without the need for
channel state information at the transmitter (CSIT) and transmit power 
control (TPC).  In this 
paper, we exploit the advantages of both SM and PRPP simultaneously. We 
propose a pseudo-random phase precoded SM (PRPP-SM) scheme, where both the 
modulation bits and the antenna index bits are precoded by pseudo-random 
phases. The proposed PRPP-SM system gives significant performance gains over 
SM system without PRPP and PRPP system without SM. Since maximum likelihood 
(ML) detection becomes exponentially complex in large dimensions, we propose 
low complexity local search based detection (LSD) algorithm suited for 
PRPP-SM systems with large precoder sizes. Our simulation results show that 
with 4 transmit antennas, 1 receive antenna, $5\times 20$ pseudo-random 
phase precoder matrix and BPSK modulation, the performance of PRPP-SM using 
ML detection is better than SM without PRPP with ML detection by about 9 dB 
at $10^{-2}$ BER. This performance advantage gets even better for large 
precoding sizes.
\end{abstract}

\vspace{-2mm}
\section{Introduction}
\label{sec1}
\vspace{-2mm}
The link reliability in single-input-single-output (SISO) fading channels 
is poor due to lack of diversity. One way to improve the link reliability 
is to get diversity gains through the use of multiple antennas. Diversity
gains can be achieved even in single-antenna systems using rotation coding 
\cite{tse} or transmit power control \cite{vinod}. Transmit power control 
requires channel state information at the transmitter (CSIT). Whereas, 
rotation coding does not require CSIT. The idea in rotation coding is to 
use multiple channel uses and precode the transmit symbol vector using a 
phase precoder matrix without requiring more slots than the number of 
symbols precoded. A $2\times 2$ phase precoder matrix with optimized 
phases is shown to achieve a diversity gain of two in SISO fading channels 
\cite{tse}. In \cite{ramesh}, the rotation coding idea has been exploited
for large precoder sizes. Instead of using optimized phases in the precoder
matrix (solving for optimum phases for large precoder sizes is difficult), 
pseudo-random phases are used. Also, the issue of detection complexity at 
the receiver for large precoder sizes has been addressed by using the 
low complexity likelihood ascent search (LAS) algorithm in \cite{lmimo1}. 
It has been shown that with pseudo-random phase precoding (PRPP) and LAS
detection, near-exponential diversity is achieved in a SISO fading channel
for large precoder sizes (e.g., $300\times 300$ precoder matrix).

Recently, spatial modulation (SM) is getting increasingly popular for
multi-antenna communications \cite{ic3},\cite{sm}. SM is a transmission 
scheme that uses multiple transmit antennas but only one transmit RF chain 
(thus requiring less RF hardware size, cost and complexity). At each time 
instant, only one among all the transmit antennas will be active and the 
others remain silent. The index of the active transmit antenna will also 
convey information bits in addition to the information bits conveyed 
through modulation symbols (e.g., QAM). An advantage of SM over 
conventional modulation is that, for a given spectral efficiency, 
conventional modulation requires a larger modulation alphabet size than
SM. For example, conventional modulation (with 1 transmit antenna and 
1 transmit RF chain) requires 8-QAM or 8-PSK to achieve 3 bpcu spectral 
efficiency. Whereas, in SM (with 4 transmit antennas and 1 transmit RF 
chain) the same spectral efficiency can be achieved using BPSK. This is
because while the BPSK symbol can convey 1 bit, 2 additional bits can be 
conveyed through the index of the chosen transmit antenna. This 
possibility of using a smaller modulation alphabet size in SM, in turn, 
results in SNR gains (for a given probability of error performance) 
in favor of SM over conventional modulation \cite{marco},\cite{tln}. 

In this paper, we exploit the advantages of both SM and PRPP simultaneously. 
Our contributions in this paper are as follows.
\vspace{-2mm}
\begin{itemize}
\item We propose a method to precode both the modulation bits and 
the antenna index bits in SM systems using pseudo-random phases. We
refer to this system as PRPP-SM system. The novelty here is that
while conventional PRPP system uses a square precoding matrix of size 
$p\times p$ (where $p$ is the number of channel uses), in PRPP-SM 
system, in order to precode the antenna index bits in addition to
the modulation bits, we use a rectangular precoding matrix of size 
$p\times pn_t$ (where $n_t$ is the number of transmit antennas).
\item
For small precoder sizes (e.g., $5\times 20$), we demonstrate using 
ML detection that the PRPP-SM system significantly outperforms PRPP 
system (without SM) and SM system (without PRPP), for the same spectral
efficiency. This performance advantage is because of the SNR gain due 
to the use of smaller alphabet size and diversity gain due to phase 
precoding.
\item
For large precoding sizes, we propose a low complexity detection
algorithm based on local search. The novelty here is a suitable
neighborhood definition that takes into account the antenna index 
bits in the PRPP-SM signal set. 
\end{itemize}

The rest of the paper is organized as follows. SM and PRPP are introduced
in Section \ref{sec2}. The proposed PRPP-SM system and the detection 
algorithm are presented in Section \ref{sec3}. Simulation results
and discussions are presented in Section \ref{sec4}. Conclusions are
presented in Section \ref{sec5}.

\begin{figure}[t]
\centering
\includegraphics[height=0.75in,width=3.4in]{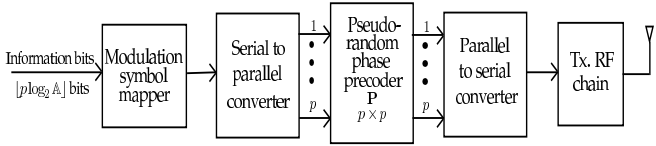}
\caption{PRPP transmitter.}
\label{fig_prpp}
\end{figure}

\section{PRPP and SM systems}
\label{sec2}
In this section, we briefly introduce PRPP and SM systems.
\subsection{PRPP system}
\label{sec2a}
\begin{figure}[t]
\centering
\includegraphics[height=2.5in,width=3.5in]{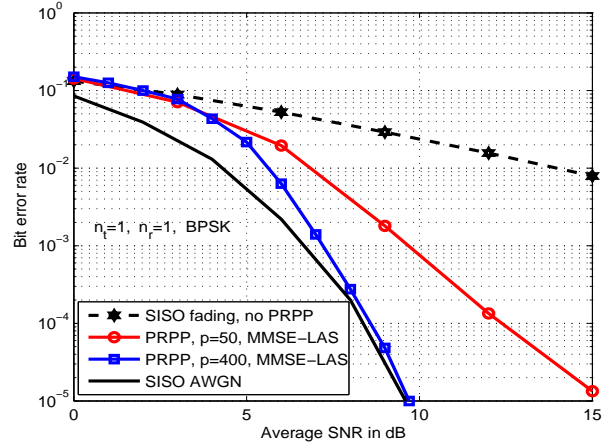}
\caption{Performance of PRPP in SISO fading with $p=50,400$, BPSK, and
LAS detection.}
\label{fig_prpp_perf}
\end{figure}

Figure \ref{fig_prpp} shows the PRPP transmitter. It takes $p$ modulated 
symbols and forms the symbol vector $\vs \in \sa^p$, where $\sa$ is the 
modulation alphabet. The symbol vector $\vs$ is then precoded using a 
$p\times p$ precoding matrix $\pp$ to get the transmit vector 
$\pp\mathbf s$. The $(r,c)$th entry of the precoder matrix $\pp$ 
is $\frac{1}{\sqrt{p}}e^{j\theta_{r,c}}$, where the phases 
$\{\theta_{r,c}\}$ are generated using a pseudo-random sequence 
generator. The seed of this random number generator is pre-shared 
among the transmitter and receiver. The precoded sequence $\pp\vs$ is 
transmitted through the channel, which is assumed to be frequency-flat 
fading. The channel fade coefficients are assumed to be i.i.d from one 
channel use to the other. At the receiver, after $p$ channel uses, the 
received symbols are accumulated to form the $p\times 1$ received
vector $\vy$, given by
\begin{eqnarray}
\vy & = &\md\pp\mathbf s + \vn \nonumber \\
& = & \mg\mathbf s + \vn,
\end{eqnarray}
where $\md=\text{diag}\{h_{(1)}\, h_{(2)}\, \cdots\, h_{(p)}\}$, 
$\mg=\md\pp$, $h_{(i)}$s are i.i.d. complex Gaussian fade coefficients 
with zero mean and unit variance, and $\vn$ is the $p\times 1$ noise 
vector $[n_{(1)}\, n_{(2)}\, \cdots\, n_{(p)}]^T$
whose entries are distributed as ${\mathbb C}{\mathcal N}(0,\sigma^2)$. 
The entries of the matrix $\mg$ are uncorrelated and 
$\lVert \md\rVert_F=\lVert \mg \rVert_F$. This creates a $p\times p$ 
virtual MIMO system. It has been shown in \cite{ramesh} that as the 
precoder size becomes large (e.g., $p=300$) the performance of PRPP 
in SISO fading, using the likelihood ascent search (LAS) detection 
algorithm in \cite{lmimo1} with MMSE initial solution, approaches 
exponential diversity performance (i.e., close to SISO AWGN performance). 
This point is illustrated in Fig.  \ref{fig_prpp_perf} which shows the 
performance of PRPP with BPSK modulation for $p=50$ and 400 in SISO
fading channels.

\subsection{SM system}
\label{sec2b}
\begin{figure}[h]
\centering
\includegraphics[height=1.1in,width=1.5in]{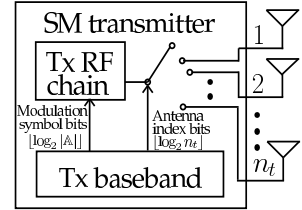}
\caption{SM transmitter with $n_t$ antennas and one transmit RF chain.}
\vspace{-2mm}
\label{sm_fig}
\end{figure}

The SM system uses $n_t$ transmit antennas but only one transmit RF chain
as shown in Fig. \ref{sm_fig}. The number of transmit RF chains, $n_{rf}=1$. 
In a given channel use, the transmitter selects one of its $n_t$ transmit 
antennas, and transmits a modulation symbol from the alphabet $\sa$ 
on the selected antenna. The number of bits transmitted per channel use 
through the modulation symbol is $\lfloor \log_2|{\mathbb A}| \rfloor$, 
and the number of bits conveyed per channel use through the index of 
the transmitting antenna is $\lfloor \log_2n_t \rfloor$. Therefore, a total 
of $\lfloor \log_2|{\mathbb A}|\rfloor + \lfloor \log_2n_t\rfloor$ 
bits per channel use (bpcu) 
is conveyed. For example, in a system with $n_t=2$, 8-QAM, the system 
throughput is 4 bpcu. 

\begin{figure*}[t]
\centering
\includegraphics[height=2in,width=7in]{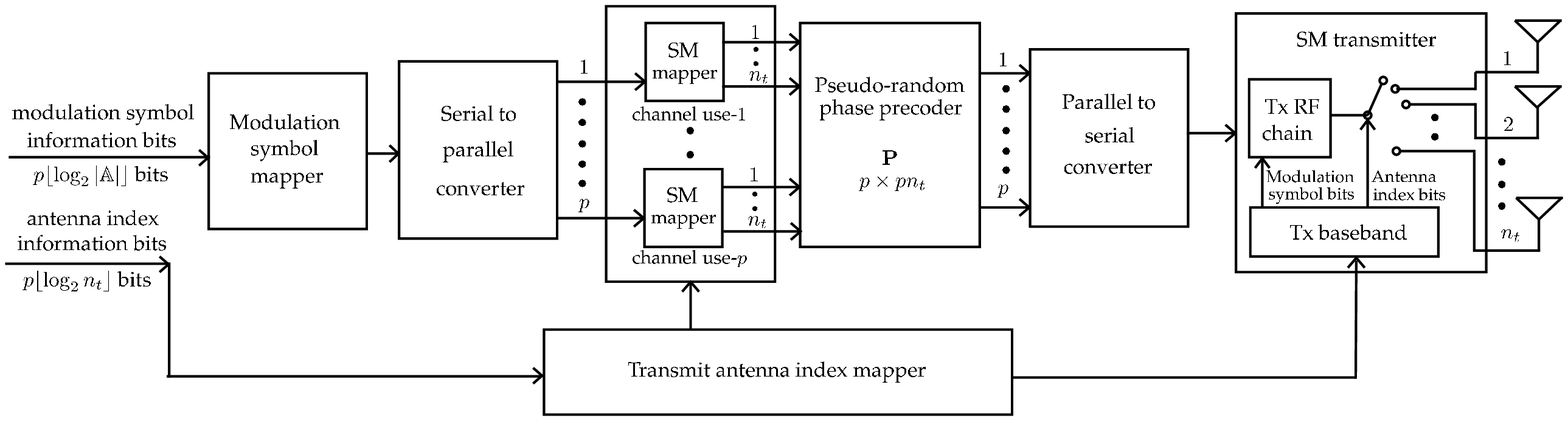}
\caption{Proposed PRPP-SM system.}
\vspace{0mm}
\label{sys}
\end{figure*}

The SM alphabet set for a fixed $n_t$ and $\sa$ is given by 
\begin{eqnarray}
\sm = 
\big \{ {\bf x}_{j,l}:j=1,\cdots,n_t, \ \ l=1,\cdots,|{\mathbb A}| \big \}, \nonumber \\ 
\mbox{s.t.} \ \ {\bf x}_{j,l} = 
[0,\cdots,0,\hspace{-4mm}\underbrace{s_{l}}_{{\scriptsize{\mbox{$j$th coordinate}}}}\hspace{-3.5mm},0,\cdots,0]^T, \ \ x_l \in \mathbb{A}.  
\label{eqy1}
\end{eqnarray}
For example, for $n_t=2$ and 4-QAM, ${\mathbb S}_{n_t,{\mathbb A}}$ is given by  
{\small
\begin{eqnarray}
\hspace{-4mm}
{\mathbb S}_{2,\mbox{{\tiny 4-QAM}}}  
\hspace{-3mm}&=&\hspace{-3mm}\Bigg\{ 
\begin{bmatrix} +1+j \\ 0 \end{bmatrix}, 
\begin{bmatrix} +1-j \\ 0 \end{bmatrix}, 
\begin{bmatrix} -1+j \\ 0 \end{bmatrix}, 
\begin{bmatrix} -1-j \\ 0 \end{bmatrix}, \nonumber \\ 
& & 
\begin{bmatrix} 0 \\ +1+j \end{bmatrix}, 
\begin{bmatrix} 0 \\ +1-j \end{bmatrix}, 
\begin{bmatrix} 0 \\ -1+j \end{bmatrix}, 
\begin{bmatrix} 0 \\ -1-j \end{bmatrix} 
\Bigg \}. 
\end{eqnarray}
}

\vspace{-4mm}
Let $\vx \in {\mathbb S}_{n_t,{\mathbb A}}$ denote the transmit vector.
Let $\mh \in \mathbb{C}^{n_r\times n_t}$ denote the channel gain matrix,
where $H_{i,j}$ denotes the channel gain from the $j$th transmit antenna 
to the $i$th receive antenna, assumed to be i.i.d complex Gaussian with
zero mean and unit variance. The received signal at the $i$th receive 
antenna is 
\begin{equation}
y_i = H_{i,j}x_l + n_i, \quad \quad i=1,\cdots,n_r,
\end{equation}
where $n_r$ is the number of receive antennas, $x_l$ is the $l$th symbol 
in ${\mathbb A}$ transmitted by the $j$th antenna, and $n_i$ is the noise
component. The signals received at all the receive antennas can be written 
in vector form as
\begin{eqnarray}
\vy& = & \mh\vx+\vn,
\label{sysmodel}
\end{eqnarray}
For this system model, the ML detection rule is given by
\begin{equation} 
\label{ml1} 
\hat{\vx}=\argmin_{\vx\in \sm} \ \|\vy-\mh\vx\|^2.  
\end{equation}

\section{Proposed PRPP-SM system}
\label{sec3}
The proposed PRPP-SM transmitter consists of $n_t$ transmit antennas and
$n_{rf}=1$ transmit RF chains as shown in Fig. \ref{sys}. It takes 
$p$ modulated symbols and forms the symbol vector $\vx_s \in \sa^p$, 
where $\sa$ is the modulation alphabet.
Let the matrix $\ma$ of size $pn_t\times p$ denote the transmit antenna 
activation pattern, such that $\ma\vx_s \in \sm^p$, where $\sm$ is the
SM signal set given by (\ref{eqy1}). The matrix $\ma$ 
consists of $p$ submatrices $\ma_{(i)}$, $i=1,\cdots,p$, each of size 
$n_t\times p$, such that 
$\ma=[\ma_{(1)}^T \, \ma_{(2)}^T \, \cdots \, \ma_{(p)}^T]^T$. The
submatrix $\ma_{(i)}$ is constructed as 
\begin{equation}
\ma_{(i)}=[{\bf 0}_{(1)} \, \cdots \, {\bf 0}_{(i-1)} \, {\bf a}_{(i)} \, {\bf 0}_{(i+1)} \, \cdots \, {\bf 0}_{(p)}],
\end{equation} 
where ${\bf 0}_{(k)}$ is a $n_t\times 1$ vector of zeroes, and 
${\bf a}_{(i)}$ is a $n_t\times 1$ vector constructed as
\begin{equation}
{\bf a}_{(i)}=[0 \, \cdots \, 0 \, \hspace{-2mm}\underbrace{1}_{\mbox{\tiny{$j_i$th coordinate}}} \hspace{-2mm} 0 \, \cdots \, 0]^T,
\end{equation}
where $j_i$ is the index of the active antenna during the $i$th 
channel use. Note that, with the above definitions, $\ma_{(i)}\vx_s \in \sm$.
For example, in a system with $n_t=2$ and $p=3$, to activate antennas
1, 2 and 1 in three consecutive channel uses, respectively, the matrix 
$\ma$ is given by
\begin{equation}
\ma \ = \ 
\begin{bmatrix}
\ma_{(1)}\\
\ma_{(2)}\\
\ma_{(3)}
\end{bmatrix}=
\begin{bmatrix}
1 & 0 & 0\\
0 & 0 & 0\\ \hdashline
0 & 0 & 0\\
0 & 1 & 0\\ \hdashline
0 & 0 & 1\\
0 & 0 & 0
\end{bmatrix}.
\label{eqx1}
\end{equation}
Note that the indices of the non-zero rows in matrix $\ma$ gives the support 
of the spatially modulated vector $\ma\vx_s\in\sm^p$. For example, in 
(\ref{eqx1}), the support given by $\ma$ is $\{1,4,5\}$.

The $\ma\vx_s$ vector is then precoded as $\pp\ma\vx_s$, using a rectangular 
precoder matrix $\pp$ of size $p\times pn_t$. The $(r,c)$th entry of the 
$\pp$ matrix is $\frac{1}{\sqrt{p}}e^{j\theta_{r,c}}$, where the 
phases $\{\theta_{r,c}\}$ are generated using a pseudo-random sequence
generator, whose seed is pre-shared among the transmitter and receiver.
The output of the precoder is transmitted on the selected antenna in 
each channel use\footnote{{\em Remark:} In the proposed PRPP-SM system,
precoding is applied to both the modulation bits as well as the antenna 
index bits, i.e., the transmit vector is $\ma\pp\ma\vx_s$. Instead, if
only the modulation bits are precoded, the transmitted vector will be 
$\ma\pp\vx_s$. When the antenna index bits are not precoded, the system 
fails to provide the diversity gain advantage of PRPP to the antenna index 
bits, and hence has a poor BER performance. The proposed PRPP-SM system, 
on the other hand, achieves very good performance because of the precoding
of the antenna index bits as well.}.

Let $n_r$ denote the number of receive antennas. The $pn_r\times 1$ 
received signal vector at the receiver is given by
\begin{eqnarray}
\vy &=&\md\ma\pp\ma\vx_s + \vn,
\end{eqnarray}
where $\md=\mbox{diag}\{\mh_{(1)}\, \mh_{(2)}\, \cdots\, \mh_{(p)}\}$, 
$\mh_{(i)}$ is the $n_r\times n_t$ channel matrix of the $i$th channel
use, the elements of $\mh_{(i)}$ are i.i.d. complex Gaussian with
zero mean and unit variance, $\vn$ is the $pn_r\times 1$ noise 
vector $[\vn_{(1)}^T\, \vn_{(2)}^T\, \cdots\, \vn_{(p)}^T]^T$ where the 
entries of $\vn_{(i)}$ are distributed as $\mathbb{C}{\cal N}(0,\sigma^2)$. 
Note  that $\lVert \md\ma\pp\ma\rVert_F=\lVert\md\ma\rVert_F$. This 
creates a $pn_r\times p$ virtual MIMO system. For this system model, 
the ML detection rule is given by
\begin{equation} 
\label{ml3} 
\{\hat{\vx}_s,\hat{\ma}\}=\argmin_{\vx_s\in \sa^p, \forall \ma} \ \|\vy_p-\md\ma\pp\ma\vx_s\|^2.  
\end{equation}
The indices of the non-zero rows in $\hat{\mathbf A}$ and the entries of 
$\hat{\mathbf x}_s$ are demapped to obtain the information bits.

Note that the ML solution in (\ref{ml3}) can be computed only for small 
precoder sizes because of its exponential complexity in $p$, i.e.,
$O((|\sa|n_t)^p)$). In Section \ref{sec4}, we will establish the 
superiority of the PRPP-SM over conventional PRPP and SM systems using 
ML detection. For large precoder sizes, we propose a low complexity 
detection algorithm in the following subsection.

\subsection{Proposed PRPP-SM detector}
\label{sec3a}
In this subsection, we propose a local search based detection (LSD)
algorithm that achieves near-ML performance in PRPP-SM systems with 
large $p$ at a low computational complexity. The local search detector 
obtains a local minima in terms of the ML cost in a local neighborhood. 
In the proposed PRPP-SM system, a key requirement for local search is a 
suitable neighborhood definition that takes into account the antenna 
index bits also. We propose the following neighborhood definition for the 
local search. The set of neighbors of a given pair of $\{\ma, \vx_s\}$, 
denoted by ${\mathcal N}(\ma, \vx_s)$, is defined as the set of all pairs 
$\{\ma', \vx_s'\}$ that satisfies one of the following three conditions:
\begin{enumerate}
\item
$\vx_s=\vx_s'$ and $\ma_{(i)}\neq\ma_{(i)}'$ for exactly a single index
$i$.
\item 
$\ma=\ma'$ and $\vx_s$ differs from $\vx_s'$ in exactly one entry.
\item
$\ma_{(i)}\neq\ma_{(i)}'$ for exactly a single index $i$, and for that 
index $i$, $x_s(i)\neq x'_s(i)$.
\end{enumerate}

For a PRPP-SM system with $n_t=2$, $p=2$, and $\sa=\{\pm1\}$, 
an example of a neighborhood is 

${\scriptsize
\hspace{1mm}
{\cal N}\left(\begin{bmatrix} 1&0 \\ 0&0 \\\hdashline 0&0 \\0&1\end{bmatrix},\begin{bmatrix} +1 \\ -1\end{bmatrix}\right)
\hspace{-1mm}=\hspace{-1mm}\left\{\begin{matrix}\vspace{2mm}
\left\{\begin{bmatrix} 1&0 \\ 0&0 \\\hdashline 0&1 \\0&0\end{bmatrix},\begin{bmatrix} +1 \\ -1\end{bmatrix}\right\},
\left\{\begin{bmatrix} 1&0 \\ 0&0 \\\hdashline 0&1 \\0&0\end{bmatrix},\begin{bmatrix} +1 \\ +1\end{bmatrix}\right\},\\
\vspace{2mm}
\left\{\begin{bmatrix} 1&0 \\ 0&0 \\\hdashline 0&0 \\0&1\end{bmatrix},\begin{bmatrix} -1 \\ -1\end{bmatrix}\right\},
\left\{\begin{bmatrix} 1&0 \\ 0&0 \\\hdashline 0&0 \\0&1\end{bmatrix},\begin{bmatrix} +1 \\ +1\end{bmatrix}\right\},\\
\left\{\begin{bmatrix} 0&0 \\ 1&0 \\\hdashline 0&0 \\0&1\end{bmatrix},\begin{bmatrix} +1 \\ -1\end{bmatrix}\right\},
\left\{\begin{bmatrix} 0&0 \\ 1&0 \\\hdashline 0&0 \\0&1\end{bmatrix},\begin{bmatrix} -1 \\ -1\end{bmatrix}\right\}
\end{matrix}
\right\}.
}$

The proposed LSD algorithm starts with an initial solution 
$\{\ma^{(0)}, \vx_s^{(0)}\}$, which is also the current solution. 
Using the defined neighborhood, the algorithm considers all the 
neighbors of $\{\ma^{(0)}, \vx_s^{(0)}\}$ and searches 
for the neighbor with the least ML cost which also has a lower ML cost 
than the current solution. If such a neighbor is found, then this neighbor 
is designated as the current solution. This marks the completion of one 
iteration of the algorithms. The iterations are repeated until a local 
minima is reached (i.e., there is no neighbor better than the current 
solution). The solution corresponding to the local minima is declared as 
the final output $\{\hat{\ma}, \hat{\vx}_s\}$. This algorithm is listed in 
{\bf Algorithm \ref{algo}} below.

\begin{algorithm}      
\caption{Listing of the proposed LSD algorithm}
\begin{algorithmic} [1] 
\STATE $\mathbf{Input: y, D}$, $\mathbf P$      
\STATE Initial solution : $\{\ma^{(0)}, \vx_s^{(0)}\}$,  $\{\hat{\ma}, \hat{\vx}_s\}=\{\ma^{(0)}, \vx_s^{(0)}\}$
\STATE Compute ${\cal N}(\hat{\ma}, \hat{\vx}_s)$
\STATE $\{\ma^c, \vx_s^c\}$ = $\argmin\limits_{\{\mb,\vz\}\in{\cal N}(\hat{\ma}, \hat{\vx}_s)} \, \|\vy-\md\mb\pp\mb\vz\|^2$
\IF{ $\|\vy-\md\ma^c\pp\ma^c\vx_s^c\|^2 < \|\vy-\md\hat{\ma}\pp\hat{\ma}\hat{\vx}_s\|^2 $}
\STATE  $\{\hat{\ma}, \hat{\vx}_s\}=\{\ma^c, \vx_s^c\}$
\STATE  Go to step 3
\ENDIF 
\STATE $\mathbf{Output}$ : $ \{\mathbf {\hat{A}},\mathbf {\hat{s}}\}$
\vspace{2mm}
\end{algorithmic}
\label{algo}
\end{algorithm} 

{\em Computing the initial support and initial solution
$\{\ma^{(0)},\vx_s^{(0)} \}$}:
The algorithm needs the initial support matrix $\ma^{(0)}$ and the 
initial solution vector $\vx_s^{(0)}$. The initial support matrix is 
obtained as follows. Obtain a $pn_t\times 1$ vector $\vv$ through an
MMSE estimator as
\[
\vv=(\md^H\md+\sigma^2\mi)^{-1}\md^H\vy. 
\]
The vector $\vv$ consists of $p$ subvectors $\vv_{(i)}$, $i=1,\cdots,p$,  
each of size $n_t\times 1$, such that
$\vv=[\vv_{(1)}^T \,\vv_{(2)}^T \, \cdots \, \vv_{(p)}^T]^T$.
The indices of the elements with the largest magnitude in each
$\vv_{(i)}$ are taken as the indices of the non-zero rows of $\ma^{(0)}$. 
The initial solution vector is obtained through an MMSE estimator as 
\[
\vx_s^{(0)} = {\cal Q}\left((\mf^H\mf+\sigma^2\mi)^{-1}\mf^H\vy\right),
\]
where $\mf=\md\ma^{(0)}\pp\ma^{(0)}$ and
${\cal Q}(.)$ denotes the Euclidean distance quantizer such that
$\vx_s^{(0)} \in \sa^p$.

\section{Simulation results}
\label{sec4}
In this section, we present the simulation results on the BER performance 
of the proposed PRPP-SM system with ML detection (for small $p$) 
and LSD detection (for large $p$). 

\begin{figure}[h]
\hspace{-4mm}
\includegraphics[height=2.75in,width=3.75in]{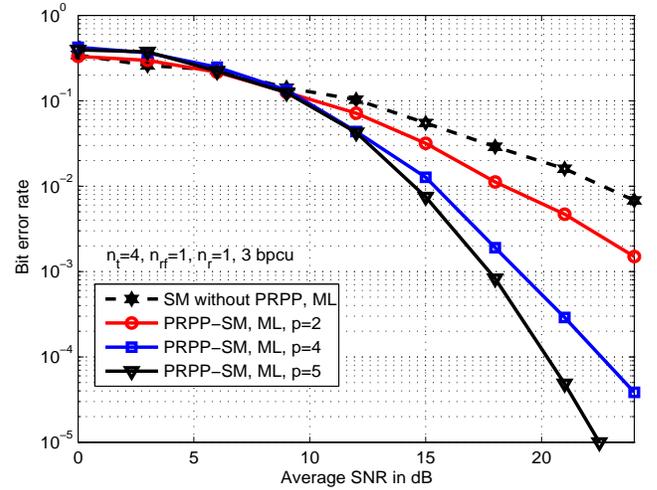}
\caption{Performance comparison between PRPP-SM system ($n_t=4, n_{rf}=1, 
n_r=1$, BPSK) with ML detection and SM system without PRPP with ML 
detection. System throughput is 3 bpcu.}
\vspace{-2mm}
\label{smml}
\end{figure}

Figure \ref{smml} compares the BER performance of PRPP-SM against the 
performance of SM without PRPP at a spectral efficiency of 3 bpcu using 
ML detection. BER plots for $n_t=4, n_{rf}=1, n_r=1$, and BPSK modulation 
for different precoder sizes $p=2,4,5$ are shown. It is observed that 
that the performance of PRPP-SM is better than SM without PRPP by about 
9 dB at $p=5$ and $10^{-2}$ BER. This performance advantage in favor of
the PRPP-SM system is due to the diversity gain offered by the 
pseudo-random phase precoding. 

\begin{figure}
\hspace{-4mm}
\includegraphics[height=2.75in,width=3.75in]{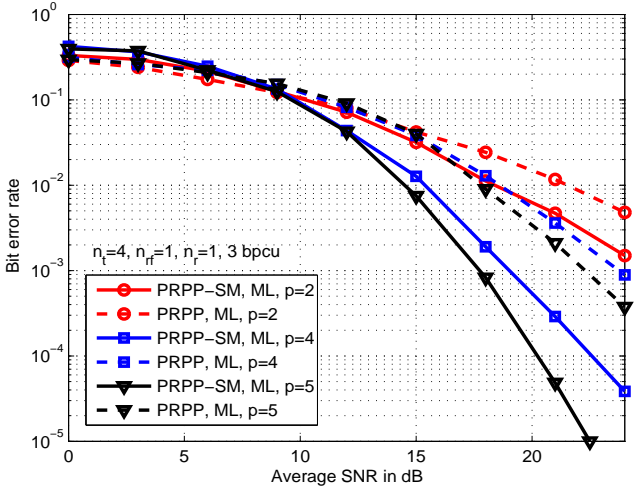}
\caption{Performance comparison between PRPP-SM system ($n_t=4, n_{rf}=1, 
n_r=1$, BPSK) with ML detection and PRPP system without SM ($n_t=1, 
n_{rf}=1, n_r=1$, 8-QAM) with ML detection. System throughput is 
3 bpcu.}
\vspace{-2mm}
\label{sisoml}
\end{figure}

Figure \ref{sisoml} compares the performance of PRPP-SM against the 
performance of PRPP without SM (i.e., $n_t=1$, $p>1$) at a spectral 
efficiency of 3 bpcu using ML detection. Here, the PRPP-SM system has 
$n_t=4, n_{rf}=1, n_r=1$, BPSK modulation, and the PRPP system without 
SM has $n_t=1, n_r=1$, 8-QAM with varying precoder sizes $p=2,4,5$. 
It is observed that the performance of PRPP-SM is better than the 
PRPP system without SM by about 4 dB at $p=5$ and $10^{-2}$ BER. 
This performance advantage in favor of PRPP-SM system is mainly 
due to the SNR gain in using BPSK in PRPP-SM against using 8-QAM  
in PRPP with out SM, for the same spectral efficiency of 3 bpcu.

\begin{figure}
\hspace{-4mm}
\includegraphics[height=2.75in,width=3.75in]{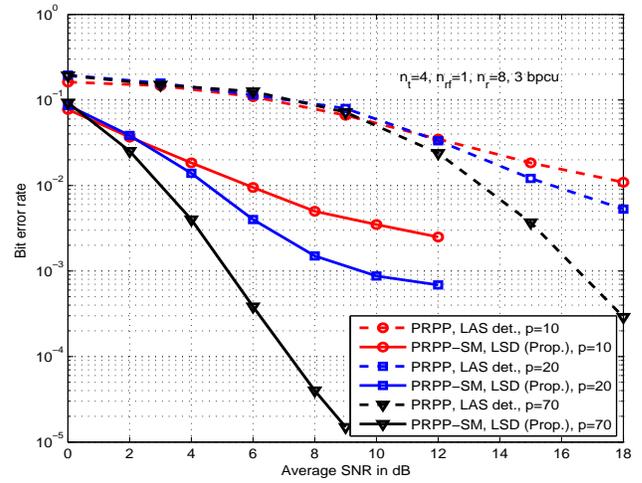}
\caption{Comparison of the BER performance of PRPP-SM ($n_t=4, n_{rf}=1, 
n_r=8$, BPSK) using the proposed LSD algorithm with that of PRPP without SM 
($n_t=n_{rf}=1,n_r=8$, 8-QAM) with LAS detection. System throughput is
3 bpcu.}
\vspace{-2mm}
\label{sisolas}
\end{figure}

Figure \ref{sisolas} compares the performance of PRPP-SM using the 
proposed LSD algorithm against the performance of PRPP without SM using 
LAS detection from \cite{lmimo1}. MMSE initial solution is used in
both LSD and LAS algorithms. For PRPP-SM, we have used 
$n_t=4, n_{rf}=1, n_r=8$, BPSK modulation. For PRPP without SM, we have 
used $n_t=1, n_r=8,$ 8-QAM. Large precoder sizes are used here; the 
precoder sizes used in both the systems are $p=10,20,70$. The spectral 
efficiency in both the systems is 3 bpcu. It is observed that the 
performance of PRPP-SM system is better than the PRPP system without SM 
by about 10 dB at $p$=70 and $10^{-2}$ BER. This indicates the potential
of PRPP-SM system to perform very well when large precoder sizes are 
employed.

\section{Conclusions}
\label{sec5}
We proposed a novel pseudo-random phase precoder based spatial modulation 
(PRPP-SM) scheme for uncoded transmissions over fading channels. In the
proposed scheme, we precoded both the modulation bits as well as the 
antenna index bits using a rectangular precoder matrix. Our simulation 
results showed that the proposed PRPP-SM system achieved diversity gains 
and SNR gains compared to conventional PRPP and SM systems. To facilitate
the detection when large precoder sizes are used in the proposed PRPP-SM
systems, we proposed a low complexity local search algorithm which used 
a neighborhood definition suited for taking into account the antenna 
index bits as well. We note that the spatial dimension in large antenna
arrays can be exploited by transmitting the $p$-length precoded vector 
through $p$ antenna elements. This would lead to a trade-off 
between the number of transmit RF chains and number of channel uses.
Such a generalized PRPP-SM scheme is an interesting topic for further
study.

\bibliographystyle{ieeetr}

\begin{thebibliography}{99}
\bibitem{tse}
D. Tse and P. Viswanath, {\em Fundamentals of Wireless Communication},
Cambridge University Press, 2005.

\bibitem{vinod}
V. Sharma, K. Premkumar and R. N. Swamy, ``Exponential diversity achieving 
spatio-temporal power allocation scheme for fading channels,'' {\em IEEE 
Trans. Info. Theory}, vol. 54, no. 1, pp. 188-208, Jan. 2008.

\bibitem{ramesh}
R. Annavajjala and P. V. Orlik, ``Achieving near exponential diversity on 
uncoded low-dimensional MIMO, multi-user and multi-carrier systems without 
transmitter CSI,'' {\em Proc. ITA 2011}, Jan. 2011.

\bibitem{lmimo1}
K. V. Vardhan, S. K. Mohammed, A. Chockalingam, and B. S. Rajan, 
``A low-complexity detector for large MIMO systems and multicarrier CDMA 
systems,'' {\em IEEE J. Sel. Areas Commun.,} vol. 26, no. 3, pp. 473-485,
Apr. 2008.

\bibitem{ic3}
R. Mesleh, H. Hass, S. Sinaovic, C. W. Ahn, ``Spatial modulation,''
{\em IEEE Trans. Veh. Tech.}, vol. 57, no. 4, pp. 2228-2241, Jul. 2008.

\bibitem{sm}
M. Di Renzo, H. Haas, A. Ghrayeb, S. Sugiura, and L. Hanzo, ``Spatial
modulation for generalized MIMO: challenges, opportunities and 
implementation,'' {\em Proceedings of the IEEE}, vol. 102, no. 1, pp. 53-55, 
Jan. 2014. 

\bibitem{marco}
N. Serafimovski1, S. Sinanovic, M. Di Renzo, and H. Haas, ``Multiple access
spatial modulation,'' {\em EURASIP J. Wireless Commun. and Networking 2012},
2012:299.

\bibitem{tln}
P. Raviteja, T. Lakshmi Narasimhan, and A. Chockalingam, ``Multiuser 
SM-MIMO versus massive MIMO: uplink performance comparison,''
available online: arXiv:1311.1291 [cs.IT] 6 Nov 2013.

\end{thebibliography}

\end{document}